\newcommand{\ppbar}  {\ensuremath{p\bar{p}} }
\newcommand{\ttbar}  {\ensuremath{t\bar{t}} }

\newcommand{\qqbar}  {\ensuremath{q\bar{q}} }

\newcommand{\mtop}    {\ensuremath{\mathrm{M}_{\mbox{top}}} }

\newcommand{\mTT}    {\ensuremath{m_{T2}}}

\newcommand{\invfb}[1]  {\ensuremath{#1~\mathrm{fb}^{-1}}}

\newcommand{\gevccnoarg}{\ensuremath{\mathrm{GeV}/c^{2}} }

\newcommand{\measStatSyst}[3]{\ensuremath{#1 \pm #2~(\mbox{stat.}) \pm
#3~(\mbox{syst.})} }

\newcommand{\measAStatSyst}[4]{\ensuremath{#1~^{+#2}_{-#3}~(\textrm{stat.}) \pm #4~(\textrm{syst.})} }

\documentclass[letterpaper]{jpconf}
\usepackage{graphicx}

\begin{document}
\title{Top quark measurement in the CDF}

\author{Hyun Su Lee (On Behalf of the CDF collaboration)}
\address{University of Chicago, Chicago, IL 60637, USA}

\begin{abstract}
We present recent top physics results in the CDF including updates of
top mass, \ttbar cross section, single top search, forward-backward
asymmetry, and the differential cross section of \ttbar. Most of
measurements utilize close to the integrated luminosity of
3~fb$^{-1}$.
\end{abstract}

\section{Introduction}
During the last decade after discovery of top
quark~\cite{r_run1topdiscCDF,r_run1topdiscD0},
top quark has been inclusively studied. By now, the mass of the top
quark has been measured to be
172.4$\pm$1.2~\gevccnoarg\cite{masscombo} which is the most
precisely measured quark mass and \ttbar pair production cross section
has been measured as less than 10~\% of uncertainties. However, many of
another top quark property have not yet been well explored due to the
limited statistics. In the ongoing data taking at Fermilab's Tevatron
proton-antiproton collider with Collider Detector at Fermilab~(CDF),
an increasing of integrated luminosity around \invfb{3} can make us to
measure the property and also discover the unexpected phenomena from
top sector.  We describes a few of the CDF's progress of top quark
measurement in the following.  

\section{Top quark measurement}
\subsection{Top quark production}
The predominant production of top quark in the Tevatron is the \ttbar
pair production. Measurement of the \ttbar production
cross section is important for test of the predictions from
perturbative QCD calculations at high transverse momentum at the level
of 10\%. Deviation from the Standard Model~(SM) predict new physics like
resonant production. 

\begin{figure}[h]
\includegraphics[width=14pc]{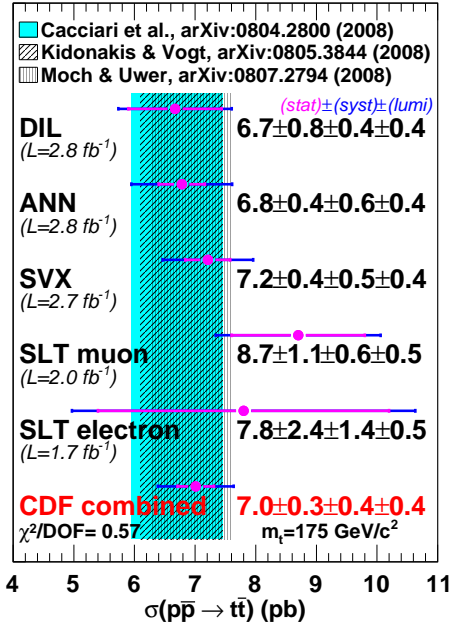}\hspace{2pc}%
\begin{minipage}[b]{14pc}\caption{\label{xsection_summ}The CDF
combination of \ttbar cross section measurement using up to \invfb{2.8}
is shown.}
\end{minipage}
\end{figure}

CDF have been measured \ttbar cross section in various different decay
topology. Fig.~\ref{xsection_summ} shows the results from various
different way in the CDF and combination with up to \invfb{2.8}.
All of results are consistent with SM prediction. 
The measured uncertainty in the combination is already reaching the relative uncertainty of the prediction
from QCD calculation.

In this combination, one of dominant uncertainty is
the luminosity measurement. This effect can be canceled taking the
ratio of the \ttbar to the $Z$ cross section by measuring the $Z$
cross section because we precisely know the $Z$ cross section from
SM calculation. Then we measure 
$$\sigma_{\ttbar}=6.9\pm0.4~(\mbox{stat.})\pm0.4(\mbox{syst.})\pm0.1(\mbox{theory})$$~\cite{cdf9474}
in the lepton jet channel which give smaller uncertainty than the CDF combination
using same amount of data.

\begin{figure}[h]
\includegraphics[width=14pc]{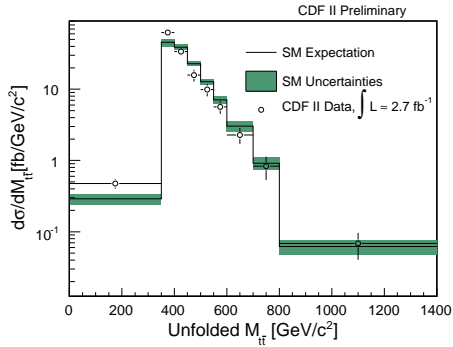}\hspace{2pc}%
\begin{minipage}[b]{14pc}\caption{\label{dsigma}
Unfolded differential cross section of \ttbar invariant mass using CDF
data is compared with SM prediction.}
\end{minipage}
\end{figure}

Since new production mechanism for top quark pairs can make the shape
of \ttbar invariant mass as resonances or general shape distortions,
the generic method to search the such contribution is to compare the
shape of the observed differential \ttbar cross
section $d\sigma/dM_{\ttbar}$ with SM expectation. The
mass of the top-antitop system is reconstructed for each event by
combining the four vectors of the four leading jets, lepton, and
missing transverse energy. The unfolding technique implemented to
correct the reconstructed distribution as for direct comparison with
theoretical differential cross section. In the update with \invfb{2.7}
data, we have in-situ jet energy scale~(JES) measurement using di-jet
mass of $W$ boson decay, which have been used in the top quark mass
measurement, that we can significantly reduce the JES systematics. As
one can see in Fig.~\ref{dsigma}, we do not find any significant difference
with SM expectation. We
check the consistency using the Anderson-Darling~(AD)
statistics~\cite{AD}. We calculate a p-value of 0.28 using AD statistics which
have a good agreement with the SM~\cite{cdfdsigma}. 

There are several searches of single top production using a variety of
multivariable techniques such as neural network, boosted decision
trees, likelihood function, and matrix element in the decay topology
with having one charge lepton~(electron 
and muon), missing transverse energy and jets~(at least one $b$-jet).
Each of measurement builds a discriminant to separate signals 
from backgrounds using multi variables. We can extract the signal from
the distribution of the discriminant. At the end, four different
discriminants can be an input of super discriminant using
neuro-evolution network. Using \invfb{2.2} of data, we measure the
single top cross section as,
$$\sigma_{st}=2.2~^{+0.7}_{-0.6}~\mbox{pb}$$~\cite{singletop}.
We have one analysis in the orthogonal decay topology which contain
missing transverse energy and jets~(at least one $b$-jet), in order to
pick up signal events which do not reconstruct electron or muon, or
contain hardronic decay tau. We build a discriminant using a neural
network technique which we can extract signal top cross section as, 
$$\sigma_{st}=4.9~^{+2.5}_{-2.1}~\mbox{pb}$$ using \invfb{2.1}
dataset~\cite{cdf9650}. 

\subsection{Top quark mass measurement}
The top mass is a fundamental parameter of the SM since it is a
dominant parameter in higher order radiative corrections to other SM
observables. The accurate determination of the top mass, combined with
precision electroweak measurement, constrains the mass of the SM Higgs
boson. 

The best precision of top quark mass by single
measurement~\cite{cdf9427} is based on
matrix element calculation of process using lepton jets channel which
we measure the top quark mass at the precision of less than 1\% level.
The combined top quark mass with \invfb{2.7} dataset is
$$\mtop=\measStatSyst{172.4}{1.0}{1.3}$$~\cite{masscombo}.
Where the uncertainty is dominated by systematic uncertainty. The CDF
have been extensive study of the systematics of top quark mass jointly
with D0, which we are now revisiting the systematics. 

There is one of interesting measurement in the dilepton channel using
$\mTT$~(transverse mass of two missing particle
system)~\cite{mT2First,mT2nd} variable which
is introduced to measure the mass of new physics 
particle in case having two missing particles. We measure the top quark
mass as, 
$$\mtop=\measAStatSyst{167.4}{4.8}{4.1}{2.9}$$
with \invfb{3.0} dataset~\cite{cdf9679}. This measurement is the first application of
the $\mTT$ variable in the real data. The methods used in this measurement will
be applicable to other measurement of new physics particles search at
the Tevatron and LHC. 

\subsection{Forward-Backward Asymmetry in Top Production}
\begin{figure}[h]
\begin{minipage}{14pc}
\includegraphics[width=14pc]{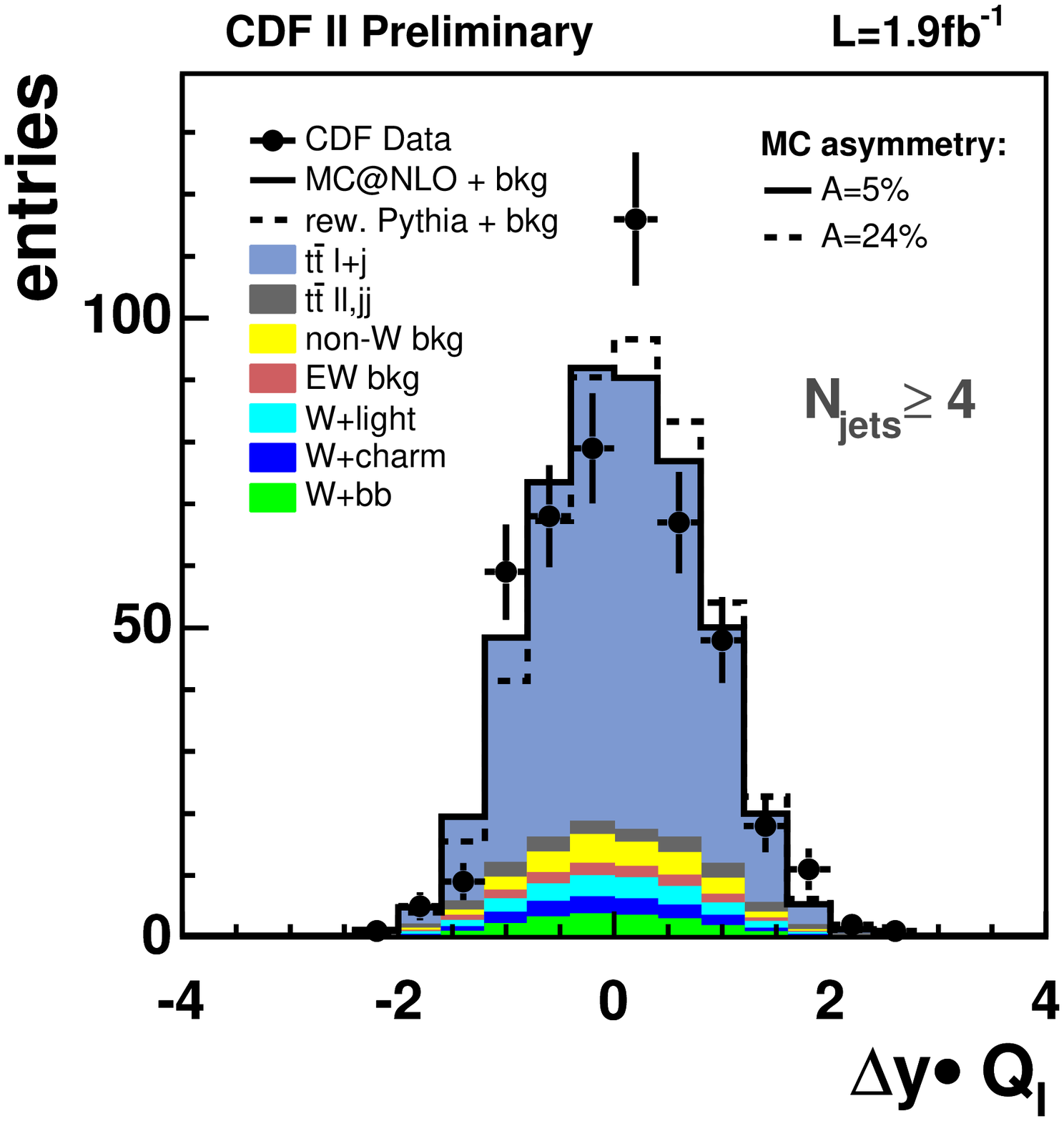}
\caption{\label{FB-ttbar}.Reconstructed rapidity difference between
top and anti-top}
\end{minipage}\hspace{2pc}%
\begin{minipage}{14pc}
\includegraphics[width=14pc]{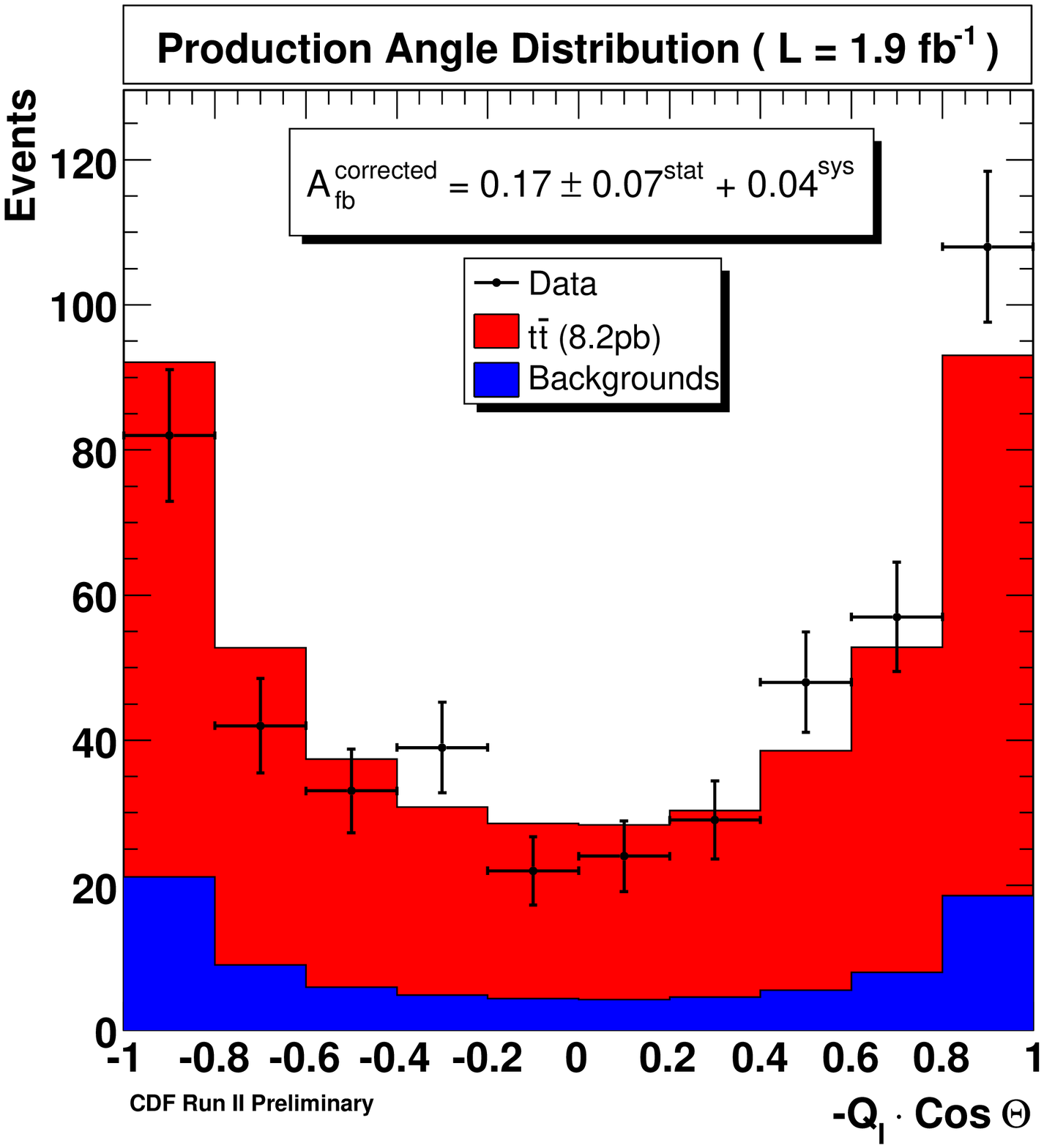}
\caption{\label{FB-ppbar}Reconstructed production angle distribution
of the top quark in Cos($\theta$).}
\end{minipage} 
\end{figure}
The forward-backward asymmetry in top production has been measured at
CDF. In the next-to-leading order~(NLO) calculation, a small charge
asymmetry, which is corresponding to a forward-backward
asymmetry~($A_{FB}$), is
calculated to be $A_{FB}=(5.0\pm1.5)\%$ in $\qqbar \rightarrow \ttbar$ .
Because top quark production in the LHC is dominated by gluon fusion,
the Tevatron is a unique place to measure this effect. CDF
have two different measurements in the different rest frame:
\ppbar~(Fig.~\ref{FB-ttbar}) and
\ttbar~(Fig.~\ref{FB-ppbar}). The measured results with \invfb{1.9} of
CDF data are
$A_{FB}^{\ppbar}~(meas)=\measStatSyst{0.17}{0.07}{0.04}$ and 
$A_{FB}^{\ttbar}~(meas)=\measStatSyst{0.24}{0.13}{0.04}$. 
This measurements
are a bit higher than the SM NLO prediction but, still it 
is consistent~\cite{forward} with SM. 

\section{Conclusions}
Number of top quark properties
not only mass and production cross section but numerous studies for
top properties have been measured. However many measurements are still
limited by the statistical uncertainty. Although we do not find
evidence conflicting with SM top
quark, we expect to have interesting
measurement with more data in near future. 
\section{Acknowledgments}
I would like to thank the CDF colleagues for their effort to carry out
these challenging physics analysis. I also thank the conference
organizers for a very rich week of physics. 

\section*{References}
\providecommand{\newblock}{}

\end{document}